\documentclass[5p,11pt,a4paper]{elsarticle}
\usepackage[english]{babel}
\usepackage{blindtext}
\usepackage[utf8]{inputenc}
\usepackage{graphicx}
\usepackage{overpic}
 \usepackage{rotating}
 \usepackage{mwe, tikz}
 \usepackage{amsmath,amssymb}
\usepackage{pgfplots}
\usetikzlibrary{calc,arrows,decorations.markings,shapes.geometric} 
\usepackage{amssymb}
\usepackage[inline]{trackchanges}
\usepackage{booktabs}
\usepackage{lipsum}
\usepackage{romannum}
\usepackage{float}
\usepackage{romannum}
\usepackage{perpage}
\MakePerPage{footnote}

\usepackage{xfp}
\usepackage{graphicx}
\usepackage{amsmath}
\usepackage{mathtools} 

\usepackage{color}
\usepackage{overpic}
\usepackage[export]{adjustbox}
\usepackage{mwe, tikz}
\usepackage{rotating}
\usepackage{latexsym}
\usepackage{csquotes}
\usepackage{tikz} 
\usetikzlibrary{calc,arrows,decorations.markings} 
\usepackage{pict2e}
\usepackage{pgfplots}
\usepackage{float}
\usepackage{hyperref}
\usepackage{xr}
\usepackage{rotating}
\usepackage{amsmath,amssymb}
\usepackage{fp}
\usetikzlibrary{calc} 
\usepackage{booktabs, tabularx}
\usepackage{romannum}
\usepackage{bm}
\usepackage{float}
\usepackage{adjustbox}
\usepackage{booktabs, tabularx}
\usepackage{longtable}
\usepackage{romannum}
\usepackage{footnote}
\usepackage{hyperref}
\usepackage[symbol*]{footmisc}

\definecolor{light-gray}{gray}{0.85}

\newcommand{\drawSlope}[6]{ 
				\coordinate (center1) at (#1,#2); 
				 \FPeval{\nx}{cos(#4*pi/180)}%
				 \FPeval{\ny}{sin(#4*pi/180)}%
				 \FPeval{\absNx}{abs(\nx)}%
				 \FPeval{\absNy}{abs(\ny)}%
				\coordinate (b) at ($ (center1) + #3*(-\ny,+\nx) $);
				\coordinate (c) at ($ (center1) - #3*(-\ny,+\nx) $); 
				 \FPeval{\xx}{(-\nx*\ny)}%
				\ifdim\xx pt < 0pt 
				\coordinate (d) at ($ (c) -2*#3*\ny*(1,0) $);
				\node at ($(d) +(0,#3*\nx)-0.2*(\nx/\absNx,0)$) {{\color{#5}$#6$}}; 
				\node at ($(d)+(#3*\ny,0)-0.2*(0,\ny/\absNy)$) {{\color{#5}$\scriptstyle 1$}}; 
				\else
				\coordinate (d) at ($ (c) +2*#3*\nx*(0,1) $);
				\node at ($(d)-#3*\nx*(0,1)-0.2*\nx/\absNx*(1,0)$) {{\color{#5}$#6$}}; 
				\node at ($(d)-#3*\ny*(1,0)-0.2*\ny/\absNy*(0,1)$) {{\color{#5}$\scriptstyle 1$}}; 
				\fi
				\draw[#5,line width=0.1mm] (d) -- (b); 
				\draw[#5,line width=0.1mm] (d) -- (c); 
				\draw[#5,line width=0.1mm] (b) -- (c); 
} 

\soulregister\ref7
\soulregister\cite7
\soulregister\citeA7
\soulregister\romannum7
\soulregister\section7

\addeditor{AN}
\pgfplotsset{compat=1.18}

\usepackage{color}

\begin{document}
\pagenumbering{arabic}

\title{Making Plasticity Visible: A Case Study for High-Temperature Exposed Aluminum Alloy}

\author[a]{Bakhtiyar Mammadli\corref{author}}
\cortext[author]{Email: \href{mailto:Bakhtiyar.Mammadli@ncbj.gov.pl}{Bakhtiyar.Mammadli@ncbj.gov.pl}}
\author[a,b]{Karol Frydrych}
\author[c]{Casim Yazici}
\author[d]{Azad Mammadli}
\author[e]{Kazimierz Skrobas}
\author[f]{Siddharth Suman}
\author[a]{F. Javier Dominguez-Gutierrez}

\address[a]{NOMATEN Centre of Excellence, National Center for Nuclear Research, ul. A. Soltana 7, 05-400 Swierk/Otwock, Poland}
\address[b]{Institute of Fundamental Technological Research, Polish Academy of Sciences, Pawińskiego 5B, 02-106 Warsaw, Poland}
\address[c]{Department of Construction, Ağrı İbrahim Çeçen University, 04400 Ağrı, Türkiye}
\address[d]{Batabat Astrophysical Observatory, Nakhchivan State University, Azerbaijan}
\address[e]{National Centre for Nuclear Research, Andrzeja Sołtana 7, Otwock, 05-400, Poland}
\address[f]{VTT Technical Research Centre of Finland, 02150 Espoo, Finland}


\begin{abstract}

Understanding the onset of plastic deformation and the early 
development of strain localization is essential for characterizing 
material behavior and identifying failure-prone regions. However, 
these transitions are not always distinguishable from conventional 
full-field strain measurements. This study investigates a data-
driven framework combining Digital Image Correlation (DIC) and 
Principal Component Analysis (PCA) to extract temporal and spatial 
deformation signatures during uniaxial tensile loading. The methodology is evaluated using AA6063-T5 aluminum alloy specimens in the as-received condition and following thermal exposure at 100~$^\circ$C, 200~$^\circ$C, and 300~$^\circ$C. PCA applied to sequential strain fields reveals 
a characteristic peak in the second principal component projection 
near the yielding transition, showing agreement with the 
conventional 0.2\% offset criterion across the investigated 
conditions. Furthermore, principal component maps extracted from 
high-temporal-resolution strain fields preceding this transition 
highlight localized deformation features associated with subsequent 
failure regions. These features are more clearly distinguishable 
than competing strain concentrations in the corresponding 
instantaneous strain maps. The results indicate that PCA-derived 
deformation signatures remain identifiable despite thermally induced 
variations in mechanical response. This approach demonstrates the 
potential of unsupervised dimensionality reduction for extracting 
physically interpretable information on plastic deformation and 
early strain localization from experimental full-field measurements.
\end{abstract}

\maketitle

\section{Introduction}
\label{sec:introduction}

The evolution of deformation and the development of strain localization during mechanical loading provide important insights into the mechanical response and failure behavior of engineering materials. Conventional mechanical testing provides valuable information about the global stress--strain response, but does not fully capture the spatial variations in deformation that may precede material failure. Full-field measurement techniques address this limitation by providing spatially resolved information on deformation, enabling the investigation of strain localization and non-uniform material behavior. Avril et al.~\cite{avril2008overview} emphasized 
that advances in digital imaging and computational processing enabled 
thousands of kinematic measurements to be obtained from a single 
experiment and stimulated the development of identification procedures 
capable of exploiting these data. More recently, Pierron and 
Grédiac~\cite{pierron2021mt20} described this change as part of a 
broader ``Material Testing 2.0'' paradigm, in which full-field 
deformation measurements and inverse identification are combined to 
extract more information from heterogeneous mechanical tests.
Among the available non-contact full-field techniques, Digital Image Correlation (DIC) has become one of the most widely used approaches in experimental mechanics. DIC derives displacement information by comparing digital images of a specimen surface during deformation and can subsequently provide spatially resolved strain fields. Early digital correlation developments were reported by Sutton et al.~\cite{sutton1983} and Peters and Ranson~\cite{peters1983}, while Chu et al.~\cite{chu1985applications} demonstrated applications of digital image correlation to experimental mechanics, including sub-pixel displacement and finite-strain measurements. The technique subsequently developed into a mature methodology for full-field deformation measurement, as reviewed by Pan et al.~\cite{pan2009}. The principal methodological advantage of DIC in the context of deformation and failure studies is not simply that strain can be measured without physical contact, but that its spatial distribution can be followed throughout loading. This distinction becomes important when deformation ceases to be approximately homogeneous and develops into localized bands, necking zones, damage regions, or cracks.

For aluminum alloys, Li et al.~\cite{Li2016DICA319} applied DIC during 
in-situ tensile testing of A319 aluminum alloy to investigate the 
relationship between local strain heterogeneity, microstructural 
features, and damage development. Their results demonstrated how full-
field measurements can reveal the progressive spatial evolution from 
heterogeneous deformation towards pronounced localization and damage. 
The investigation of spatially resolved deformation in aluminum alloys 
has also been extended to elevated-temperature conditions. Chen et 
al.~\cite{Chen2024ImprovedDIC} investigated full-field strain 
measurements in AA6082 aluminum alloy during uniaxial and biaxial 
tensile testing under hot--stamping conditions. Their experiments 
demonstrated that DIC can follow the development of localized 
deformation towards fracture under elevated temperature and large deformation conditions. These thermo--mechanical conditions 
differ from thermal pre--exposure followed by room-temperature testing, 
the results demonstrate the applicability of spatially resolved strain 
measurements to aluminum alloys exhibiting strongly evolving and 
localized deformation.

While full-field measurements provide detailed information about the spatial development of deformation, a sequence of strain maps contains a large amount of spatial and temporal information. During the transition from elastic to plastic deformation, multiple heterogeneous regions may evolve simultaneously, making characteristic changes in the deformation process difficult to quantify from individual strain maps alone. This motivates the use of materials informatics and data-driven approaches for extracting compact and physically interpretable information from evolving DIC strain fields \cite{SUMAN2021124022}. Frydrych et al.~\cite{frydrych2021materials} reviewed developments in materials informatics specifically in the context of mechanical deformation and highlighted the large amount of information contained in image-based and DIC--derived deformation data. 

An important direction is therefore the application of statistical and 
machine-learning methods to full-field measurements after the 
displacement or strain fields have been obtained \cite{Siddharth1, Prakash2021}. A sequence of DIC 
strain maps can be regarded as a high-dimensional spatio--temporal 
dataset in which each loading stage contains information about the 
spatial distribution of deformation. Unsupervised dimensionality-
reduction techniques, particularly Principal Component Analysis (PCA), 
provide a means of extracting dominant patterns from these evolving 
fields without requiring predefined labels for individual deformation 
states. Analyzing the evolution of these patterns therefore offers the 
possibility of identifying characteristic deformation transitions and 
spatial localization features without reducing the experiment solely to 
a global stress--strain response. The concept was subsequently 
demonstrated experimentally by Mäkinen et 
al.~\cite{makinen2022detection}, who combined DIC measurements with PCA 
to detect the onset of yielding in metallic tensile specimens. 
Consecutive local strain fields were treated as high-dimensional input 
vectors and decomposed into principal components, after which the 
temporal evolution of the corresponding projections was analyzed. In 
particular, the second principal-component projection exhibited a 
characteristic peak associated with the elastic-to-plastic transition. 
The approach was demonstrated experimentally for Haynes 230 and 
Hastelloy X alloys, showing that the onset of yielding can be 
identified directly from the evolution of experimentally measured local 
strain fields.
A related application of dimensionality reduction to experimental DIC measurements was demonstrated by Mammadli et al.~\cite{mammadli2025universal} for creep deformation and failure. In that study, the temporal evolution of dimensionality--reduced strain fields was analyzed to identify characteristic deformation transitions, while the corresponding spatial component maps were investigated for their ability to identify regions associated with subsequent failure development. 

For precipitation-hardened aluminum alloys, material state can be strongly influenced by thermal history. In Al--Mg--Si alloys such as AA6063, precipitation hardening plays a central role in determining strength, strain-hardening behavior, and ductility. Consequently, thermal exposure can modify the subsequent mechanical response through changes in the precipitation state. Aydi et al.~\cite{Aydi2015AA6063T5} investigated the mechanical properties and microstructure of AA6063 following extrusion and T5 heat treatment, demonstrating the close relationship between precipitation and the resulting mechanical properties. Khadyko et al.~\cite{Khadyko2017AA6063} examined AA6063 under different heat-treatment conditions and showed that temper condition influences its plastic anisotropy and mechanical response. In a subsequent study, Khadyko et al.~\cite{Khadyko2019DuctilityAA6063} investigated the tensile ductility of AA6063 in different tempers and demonstrated substantial differences in strength and ductility depending on the material condition. The evolution of precipitation during ageing and its relationship with yield strength was further investigated by Sekhar et al.~\cite{Sekhar2019AgingAA6063}. These studies establish that changes in the thermal condition of AA6063 can significantly modify its subsequent plastic deformation behavior.

The present study investigates the ability of Principal Component Analysis (PCA) to extract mechanically meaningful deformation signatures from Digital Image Correlation (DIC) measurements obtained during uniaxial tensile testing of AA6063-T5 aluminum alloy under different thermal exposure conditions. The contribution of this work is twofold. First, the temporal evolution of PCA projections is examined to assess whether the characteristic signature associated with the yielding transition remains identifiable across thermally induced variations in mechanical response. Second, spatial PCA component maps extracted from strain fields preceding this transition are analyzed to determine whether they reveal early deformation patterns associated with regions of subsequent strain localization and failure. By integrating these two analyses, the study evaluates the capability of PCA to characterize both deformation transitions and the spatial development of localization, extending its application to materials with different thermal histories.

\section{Methods}
\label{sec:methods}

\begin{table}[b!]
    \centering
    \caption{Chemical composition of AA6063-T5 aluminum alloy (wt.\%).}
    \label{tab:chemical_composition}
    \begin{tabular}{lcc}
        \hline
        Element & AA6063-T5 (wt.\%)  \\
        \hline
        Ag & $<0.001$   \\
        Al & 98.689     \\
        B  & 0.0027     \\
        Be & $<0.0003$  \\
        C  & --         \\
        Ca & 0.003      \\
        Cd & $<0.001$   \\
        Co & $<0.001$   \\
        Cr & 0.0115     \\
        Cu & 0.0486     \\
        Fe & 0.2642     \\
        Mg & 0.396      \\
        Mn & 0.0459     \\
        Mo & --         \\
        N  & --         \\
        Ni & $<0.003$   \\
        P  & --         \\
        Pb & 0.0061     \\
        S  & --         \\
        Si & 0.4147     \\
        Sn & $<0.003$   \\
        Sr & $<0.0005$  \\
        Ti & 0.0189     \\
        V  & 0.0121     \\
        Zn & 0.0778     \\
        \hline
    \end{tabular}
\end{table}

\subsection{Experimental Methodology}
\label{sec:experimental_methodology}

\begin{figure}[b!]
    \centering
    \includegraphics[width=0.985\linewidth]{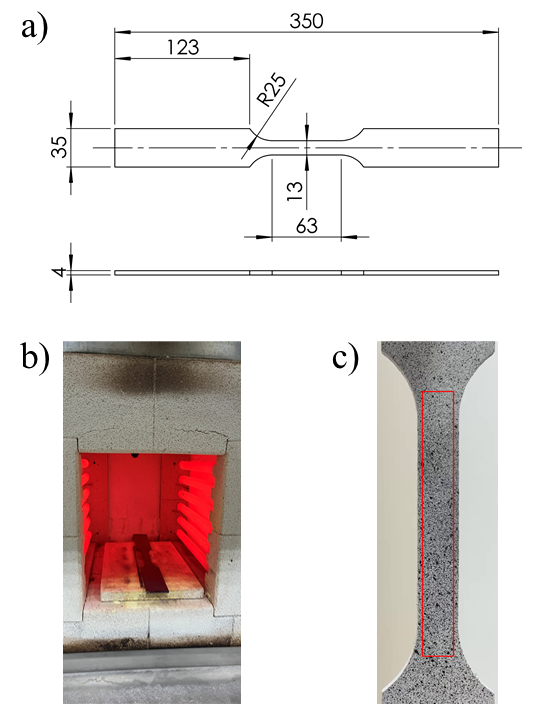}
    \caption{Experimental setup adopted in this study. (a) Geometry and principal dimensions of the AA6063-T5 aluminum tensile specimen. All dimensions are given in millimeters. (b) Laboratory furnace used for the thermal treatments. (c) Representative tensile specimen prepared for DIC; the red rectangle indicates the region of interest used for the full-field strain analysis.} 
    \label{fig:method_exp}
\end{figure}

 Flat tensile specimens were prepared from commercial AA6063--T5 aluminum alloy. The chemical compositions of both materials are presented in Table~\ref{tab:chemical_composition}. The experimental setup adopted in this study is summarized in Figure~\ref{fig:method_exp}. Similar thermal exposure and tensile testing procedures were employed in our previous investigations \cite{Casim1,Ozkal2026}. The 
 specimens were machined in accordance with ASTM E8/E8M 
 \cite{ASTM_E8_2022}, and their geometry is presented in 
 Figure~\ref{fig:method_exp}(a). The material had identical in-
 plane specimen dimensions, with an overall length of 350~mm, a grip 
 section width of 35~mm, and a reduced section width of 13~mm 
 extending over a parallel length of 63~mm. The transition between 
 the reduced and grip sections was formed with a fillet radius of 
 25~mm. The nominal specimen thicknesses were 4~mm for AA6063-T5 
 aluminum alloy. Prior to testing, the 
 actual thickness of each specimen was measured and used to 
 calculate the engineering stress. The machined edges were ground 
 and deburred, and the specimens were prepared with the extrusion or 
 rolling direction aligned with the tensile loading axis.

To investigate the influence of thermal exposure on the deformation 
behavior, the specimens were tested in five 
conditions: the untreated reference condition and specimens 
thermally exposed to 100~$^\circ$C, 200~$^\circ$C, and 300~$^\circ$C. Prior to 
mechanical testing, the specimens were subjected to thermal 
treatment using the temperature--time schedule  \cite{YAZICI2025104242}. The furnace temperature was increased from 
room temperature to the target temperature at a constant heating 
rate of 20$^\circ$C/min, followed by an isothermal holding period of 
30~min to promote a uniform temperature distribution throughout the 
specimen cross-section. After thermal exposure, the specimens were 
removed from the furnace and allowed to cool naturally in still air 
to room temperature, representing post-fire air-cooling conditions 
\cite{ZEYBEK2026149761}. All heat treatments were carried out using 
the electric laboratory furnace shown in Figure~\ref{fig:method_exp}
(b).

Quasi-static tensile tests were performed at room temperature using 
an INSTRON~5982 universal testing machine operating under 
displacement control. The machine cross-head was driven at a 
constant displacement rate of 2~mm/min in accordance with 
ASTM E8/E8M \cite{ASTM_E8_2022}. The specimens were secured using 
wedge-type mechanical grips, with particular attention paid to 
maintaining alignment between the specimen and loading axes to 
minimize unintended bending during testing. For each material and 
thermal exposure condition, one specimen was tested. During each 
experiment, the applied force and cross-head displacement were 
continuously recorded to obtain the engineering stress--strain 
response.

For Digital Image Correlation (DIC) analysis, one surface of each 
specimen was prepared by applying a uniform white matte coating 
followed by a random black speckle pattern to obtain a high-contrast 
surface suitable for image correlation. A representative specimen 
prepared for DIC analysis is shown in Figure~\ref{fig:method_exp}
(c). This surface preparation was particularly important for the 
thermally exposed specimens, as changes in surface appearance caused 
by thermal exposure could reduce image contrast and affect the 
accuracy of the correlation analysis. During tensile loading, the 
deformation of the specimen surface was continuously recorded using 
a digital camera operating at 60~frames/s. The recorded videos were 
subsequently decomposed into individual image frames and 
synchronized with the corresponding mechanical data for full-field 
strain analysis.


\subsection{Computational Methodology}
\label{sec:computational_methodology}

\begin{figure*}[htbp]
    \centering
    \includegraphics[width=0.85\linewidth]{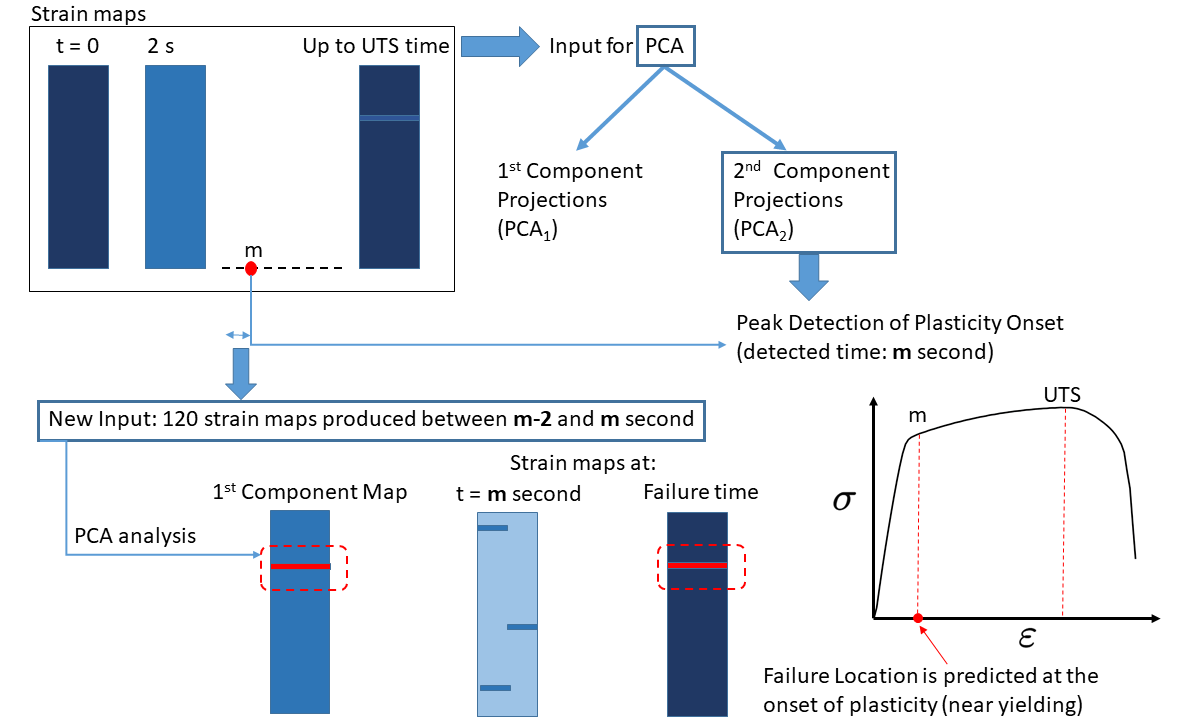}
    \caption{Overview of the proposed two-stage computational workflow. Full-field $E_{yy}$ strain maps obtained from DIC were first analyzed using PCA, where the evolution of the principal component projections was used to identify the onset of localized plastic deformation. Subsequently, PCA was performed on a high-temporal-resolution subset of strain maps to generate principal component maps for predicting the future failure location \cite{mammadli2025universal}.}
    \label{fig:method_comp}
\end{figure*}

The full-field strain measurements used in this study were obtained using the subset-based Digital Image Correlation Engine (DICe) \cite{turner2016local,turner2015dice}. In this formulation, the displacement field is determined by tracking the motion of randomly textured subsets between reference and deformed images through minimization of the image intensity residual. The optimal kinematic parameters are obtained by solving

\begin{equation}
    \mathbf{p}^{*}
    =
    \arg\min_{\mathbf{p}}
    \frac{1}{2}
    \int_{\Omega}
    \left[
    \phi(\mathbf{w}(\mathbf{x},\mathbf{p}))
    -
    \phi_{0}(\mathbf{x})
    \right]^2
    d\mathbf{x},
    \label{eq:dic_optimization}
\end{equation}

\noindent where $\Omega$ denotes the subset domain, $\mathbf{w}(\mathbf{x},\mathbf{p})$ is the mapping from reference coordinates $\mathbf{x}$ to deformed coordinates parameterized by $\mathbf{p}$, and $\phi_0$ and $\phi$ represent the normalized intensities of the reference and deformed images. The resulting displacement field was subsequently post-processed using the Virtual Strain Gauge (VSG) formulation \cite{turner2018vsg} to compute the full-field Green--Lagrange strain tensor. In the present study, only the normal strain component $E_{yy}$, corresponding to the loading direction, was used for the subsequent computational analysis.

Principal Component Analysis (PCA) was employed to identify the dominant deformation patterns within the full-field strain data, following the methodology previously reported in \cite{makinen2022detection, mammadli2025universal}. The input dataset consisted of the Green--Lagrange normal strain component $E_{yy}$ extracted from each DIC strain field. Each strain map was reshaped into a one-dimensional vector and assembled into a data matrix $X$, with each row corresponding to a single time step. Prior to PCA, the strain fields were normalized according to

\begin{equation}
    X_{(k,i)} =
    \frac{
    \left|
    E_{yy}^{(k,i)} -
    \left\langle E_{yy}^{(k,i)} \right\rangle
    \right|
    }{
    \sqrt{
    \left\langle \left(E_{yy}^{(k,i)}\right)^2 \right\rangle
    -
    \left\langle E_{yy}^{(k,i)} \right\rangle^2
    }
    },
    \label{eq:strain_normalization}
\end{equation}

\noindent where $k$ denotes the time step, $i$ denotes the spatial point, and $\langle\cdot\rangle$ represents spatial averaging over each strain field. This preprocessing centers and standardizes each strain field before taking the absolute value, thereby emphasizing the magnitude of local strain deviations and facilitating the identification of dominant deformation patterns. The resulting data matrix was subsequently decomposed using Singular Value Decomposition (SVD),

\begin{equation}
    X = U \Sigma W^{\mathrm{T}},
    \label{eq:pca_svd}
\end{equation}

\noindent where $U$ and $W$ contain the left and right singular vectors, respectively, while $\Sigma$ is a diagonal matrix containing the singular values. The squared singular values are proportional to the variance contributions of the corresponding modes in the uncentered second-moment decomposition of $X$. The temporal evolution of these deformation modes was quantified through the principal component projections,

\begin{equation}
    \mathrm{PCA}_j^k =
    \frac{\mathbf{s}_j \cdot X_k}{\sqrt{\sigma_j}},
    \label{eq:pca_projection}
\end{equation}

\noindent where $X_k$ denotes the normalized strain map at the $k$-th time step, $s_j$ is the $j$-th principal vector, and $\sigma_j$ 
is the corresponding singular value. The principal vectors were 
subsequently reshaped to the original dimensions of the strain 
fields to obtain the corresponding principal component maps, 
representing the spatial distribution of each deformation mode over 
the specimen surface.

The two-stage computational framework adopted in this study is summarized in Figure~\ref{fig:method_comp}. The principal component projections describe the temporal evolution of the dominant deformation modes throughout the tensile test, whereas the corresponding principal component maps represent their spatial distribution over the specimen surface. During the first stage, PCA was performed using the complete sequence of full-field $E_{yy}$ strain maps obtained from the beginning of the tensile test up to the Ultimate Tensile Strength (UTS). Since DIC analysis is computationally considerably more demanding than the subsequent PCA calculations, strain fields were initially evaluated at 2~s intervals. This provided sufficient temporal resolution to capture the overall deformation evolution while reducing computational cost. By restricting high-temporal-resolution DIC processing to a short interval around the identified transition, the proposed coarse-to-fine strategy substantially reduces the number of strain-field calculations compared with processing the complete deformation history at the original image acquisition rate. This approach enabled the computational analysis to be performed on a single CPU without requiring high-performance computing resources.

Among the principal component projections, the second component exhibited a distinct peak during tensile loading. The time corresponding to this peak, denoted by $m$, was used to identify the characteristic transition associated with the onset of localized plastic deformation. Once the transition time had been identified, the original tensile-test video was revisited to recover the complete image sequence recorded during the 2~s interval immediately preceding $m$. The DIC analysis was then repeated using the same processing parameters as in the initial analysis, including the subset size, step size, region of interest, and correlation parameters. Thus, only the temporal resolution was increased while maintaining identical processing conditions. This procedure generated a high-temporal-resolution subset consisting of 120 strain maps between $m-2$~s and $m$.

Finally, PCA was applied to this refined dataset to obtain principal component maps from the high-temporal-resolution strain fields. This analysis was intended to enhance the identification of subtle spatial deformation patterns immediately preceding the detected transition and to identify regions associated with the eventual failure location before localization became visually distinguishable in the original strain fields. The PCA-detected transition was used to select the analysis window in order to demonstrate that both yielding detection and identification of failure-prone regions can be achieved solely from DIC strain fields. However, the spatial PCA analysis itself does not inherently require PCA-based detection of the yielding transition, as the analysis window could alternatively be selected using the conventional yielding point determined from the stress--strain response.

\section{Results}
\label{sec:results}

\subsection{Evolution of Full-Field Strain Maps}
\label{sec:strain_evolution}

The evolution of the full-field deformation behavior was first investigated using the Green--Lagrange strain tensor components obtained from DIC. Representative strain maps at selected stages of deformation were analyzed to characterize the spatial development of deformation, identify the dominant strain component, and establish the basis for the subsequent Principal Component Analysis (PCA).

\begin{figure}[t!]
    \centering
    \includegraphics[width=\linewidth]{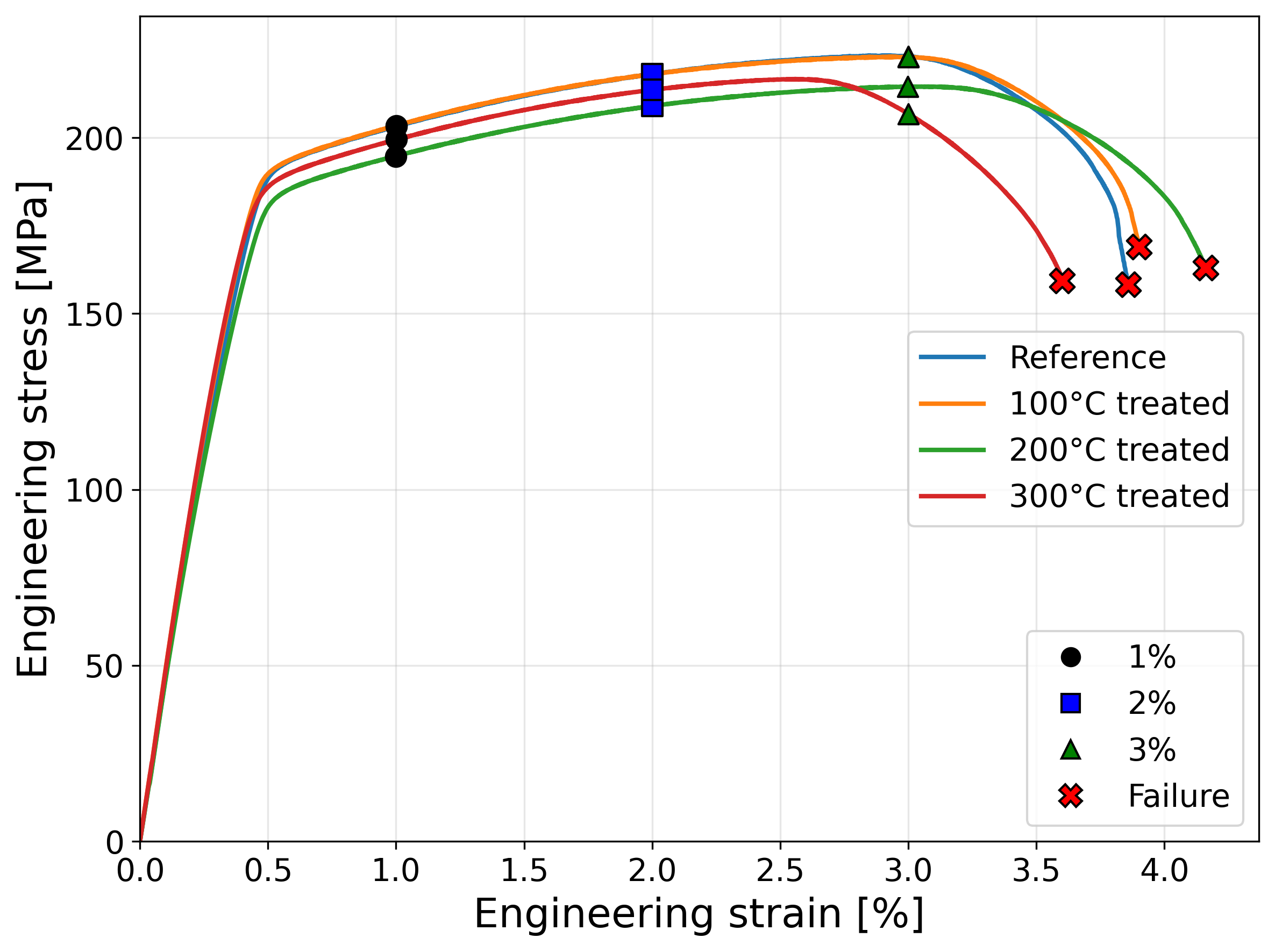}
    \caption{Engineering stress--strain curves for the reference and thermally exposed AA6063-T5 aluminum alloy specimens. The markers indicate the selected engineering strain levels of approximately 1\%, 2\%, and 3\%, together with the final state immediately preceding failure, used to display the representative full-field strain maps in Figure~\ref{fig:results_1_b}.}
    \label{fig:results_1_a}
\end{figure}

\begin{figure*}[t]
    \centering
    \includegraphics[
        width=0.95\linewidth,
        height=0.85\textheight,
        keepaspectratio
    ]{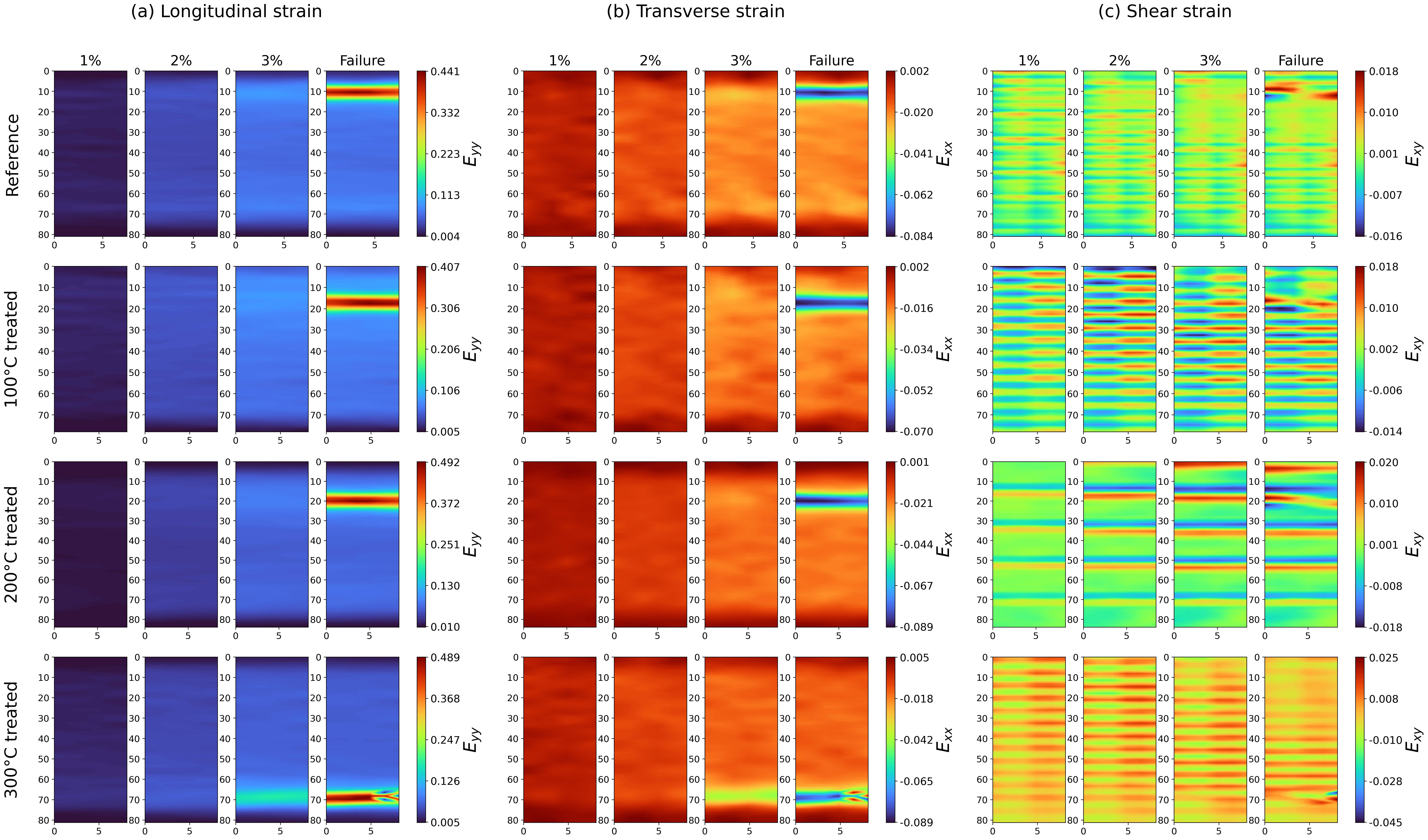}
    \caption{Evolution of the full-field Green--Lagrange strain components at selected deformation stages for the reference and thermally exposed AA6063-T5 aluminum alloy specimens: (a) longitudinal strain $E_{yy}$, (b) transverse strain $E_{xx}$, and (c) shear strain $E_{xy}$. The strain maps correspond to engineering strain levels of approximately 1\%, 2\%, and 3\%, together with the final state immediately preceding failure. The dimensions of the strain maps are given in mm.}
    \label{fig:results_1_b}
\end{figure*}

Figure~\ref{fig:results_1_a} presents the engineering stress--strain curves of the reference and thermally exposed AA6063-T5 specimens. Engineering strain levels of approximately 1\%, 2\%, and 3\%, together with the final state immediately preceding failure, were selected as common reference points for comparing the evolution of the full-field strain distributions shown in Figure~\ref{fig:results_1_b}.

The engineering stress--strain curves exhibit a typical elastic--plastic response for all specimens, with yielding followed by strain hardening up to the Ultimate Tensile Strength (UTS) and a subsequent decrease in engineering stress associated with localized deformation. The reference and 100$^\circ$C specimens exhibit similar mechanical responses, while the specimen exposed to 200$^\circ$C reaches a larger engineering strain before failure. The most pronounced change is observed for the specimen exposed to 300$^\circ$C, which exhibits substantially earlier strain localization and failure.

The longitudinal Green--Lagrange strain component $E_{yy}$ shown in Figure~\ref{fig:results_1_b}(a) illustrates the progressive development of deformation localization during tensile loading. At engineering strains of 1\% and 2\%, the strain fields remain comparatively homogeneous over the gauge section for all conditions, with only small spatial variations. As deformation increases towards 3\%, regions of elevated longitudinal strain become more apparent in the thermally exposed specimens. This effect is particularly pronounced for the specimen exposed to 300$^\circ$C, where a localized region develops earlier than in the remaining conditions. Immediately before failure, all specimens exhibit a strong concentration of $E_{yy}$ at the final fracture location.

A similar evolution is observed for the transverse strain component $E_{xx}$ shown in Figure~\ref{fig:results_1_b}(b). During the early stages of deformation, the transverse strain field remains relatively uniform, whereas increasing tensile deformation produces progressively localized contraction within the gauge section. The regions of strongest transverse contraction generally coincide with regions of elevated longitudinal strain, reflecting the coupled development of longitudinal extension and transverse contraction during plastic deformation.

The shear strain component $E_{xy}$ shown in Figure~\ref{fig:results_1_b}(c) exhibits a different spatial pattern. Alternating positive and negative shear bands are visible throughout the gauge section, becoming more pronounced as deformation progresses. The banded structure is particularly evident in the thermally exposed specimens and becomes increasingly concentrated near the fracture region during the final stage of deformation.

Among the three strain components, $E_{yy}$ reaches the largest strain magnitudes, as indicated by the corresponding color-bar ranges in Figure~\ref{fig:results_1_b}(a). Furthermore, the evolution from relatively homogeneous deformation to pronounced strain localization is most clearly represented by the longitudinal strain field. Therefore, the subsequent PCA was performed using $E_{yy}$ as the input strain component.

\begin{figure*}[htbp]
    \centering
    \includegraphics[
        width=0.95\linewidth,
        height=0.85\textheight,
        keepaspectratio
    ]{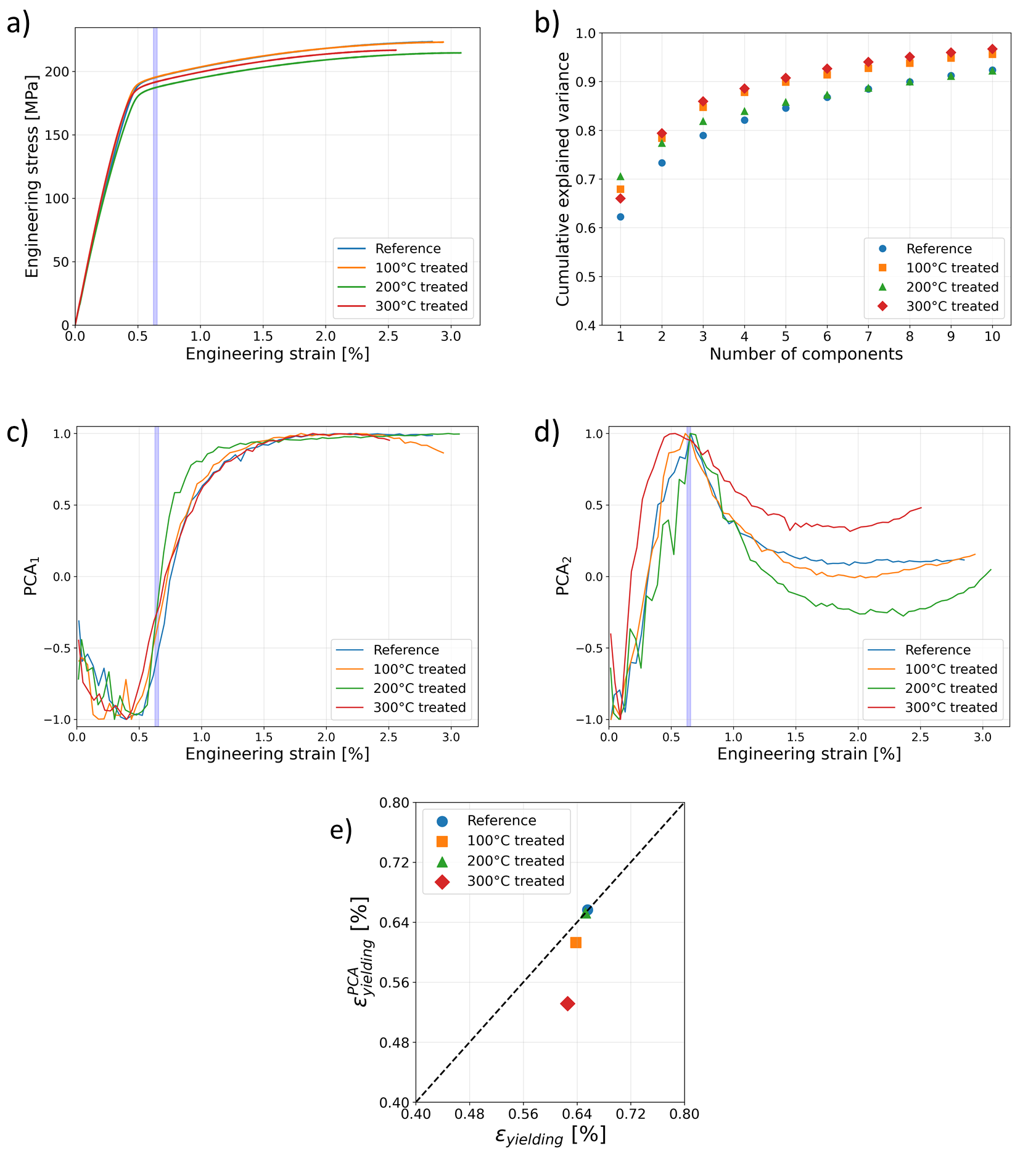}
    \caption{PCA-based identification of the yielding transition from full-field DIC strain maps. (a) Engineering stress--strain curves up to UTS for the reference and thermally exposed aluminum alloy specimens. The blue shaded region indicates the range of yield strains determined using the 0.2\% offset method. (b) Cumulative explained variance of the first 10 PCA components obtained from sequential $E_{yy}$ strain maps, showing that the first two components capture most of the dominant strain-field variation. (c) First PCA projection as a function of engineering strain, exhibiting a transition from low to high values near the yielding region. (d) Second PCA projection as a function of engineering strain, exhibiting a distinct peak close to the yielding transition. (e) Yield strain estimated from the second PCA projection, $\varepsilon_{\mathrm{yielding}}^{\mathrm{PCA}}$, compared with the yield strain determined from the engineering stress--strain curves, $\varepsilon_{\mathrm{yielding}}$. The dashed line represents the $x=y$ relation.}
    \label{fig:projections}
\end{figure*}

\subsection{Detection of the Onset of Plastic Deformation}
\label{sec:plasticity_onset}

The next stage of the analysis investigates whether the onset of plastic deformation can be identified directly from the temporal evolution of the full-field $E_{yy}$ strain maps. PCA was applied to the sequence of strain maps obtained from the beginning of the tensile test up to UTS, and the resulting principal component projections were analyzed to identify characteristic transitions in the deformation behavior.

Figure~\ref{fig:projections} presents the PCA results obtained from the full-field $E_{yy}$ strain maps. For comparison, Figure~\ref{fig:projections}(a) shows the engineering stress--strain curves up to UTS, where the blue shaded region indicates the range of yield strains determined using the conventional 0.2\% offset method.

Before analyzing the principal component projections, the cumulative explained variance was evaluated to determine the number of components required to represent the dominant deformation behavior. As shown in Figure~\ref{fig:projections}(b), the first two principal components account for approximately 80\% of the cumulative variance for all thermal exposure conditions. This indicates that most of the variation in the full-field strain evolution is represented by the first two dominant deformation modes, while higher-order components contribute comparatively less.

The temporal evolution of the first and second principal component projections is presented in Figures~\ref{fig:projections}(c) and~\ref{fig:projections}(d), respectively. The first principal component projection exhibits a smooth transition from low to high values as deformation progresses, reflecting the continuous evolution of the dominant deformation mode during tensile loading. In contrast, the second principal component projection develops a pronounced peak within the engineering strain interval highlighted by the yield region in Figure~\ref{fig:projections}(a). This characteristic behavior is consistently observed for all investigated thermal exposure conditions, indicating a strong correspondence between the evolution of the second principal component and the yielding transition.

The PCA-detected yielding strain was defined as the engineering strain corresponding to the maximum of the second principal component projection,

\begin{equation}
    \varepsilon_{\mathrm{yielding}}^{\mathrm{PCA}}
    =
    \underset{\varepsilon}{\arg\max}
    \;
    \mathrm{PCA}_{2}(\varepsilon),
    \label{eq:pca_yielding}
\end{equation}

\noindent where $\varepsilon$ denotes the engineering strain. The corresponding time was denoted by $m$ in the computational methodology.

Figure~\ref{fig:projections}(e) compares the PCA-based yielding strain, $\varepsilon_{\mathrm{yielding}}^{\mathrm{PCA}}$, with the yield strain, $\varepsilon_{\mathrm{yielding}}$, determined from the engineering stress--strain curves using the conventional 0.2\% offset method. The reference, 100$^\circ$C, and 200$^\circ$C conditions lie close to the $x=y$ relationship, indicating good agreement between the two approaches. A slightly larger deviation is observed for the specimen exposed to 300$^\circ$C. Nevertheless, the overall correspondence demonstrates that the peak of the second principal component projection provides an effective indicator of the yielding transition identified from the global mechanical response.

\subsection{Prediction of Failure Location}
\label{sec:failure_prediction}

Following the identification of the yielding transition, the final stage of the analysis investigates whether the future failure location can be identified using only a short sequence of strain maps preceding the detected transition. PCA was therefore applied to a high-temporal-resolution subset of $E_{yy}$ strain maps extracted immediately before the detected PCA$_2$ peak. The resulting principal component maps were compared with the corresponding strain maps at the detected transition and immediately before failure.

\begin{figure*}[htbp]
    \centering
    \includegraphics[
        width=0.95\linewidth,
        height=0.85\textheight,
        keepaspectratio
    ]{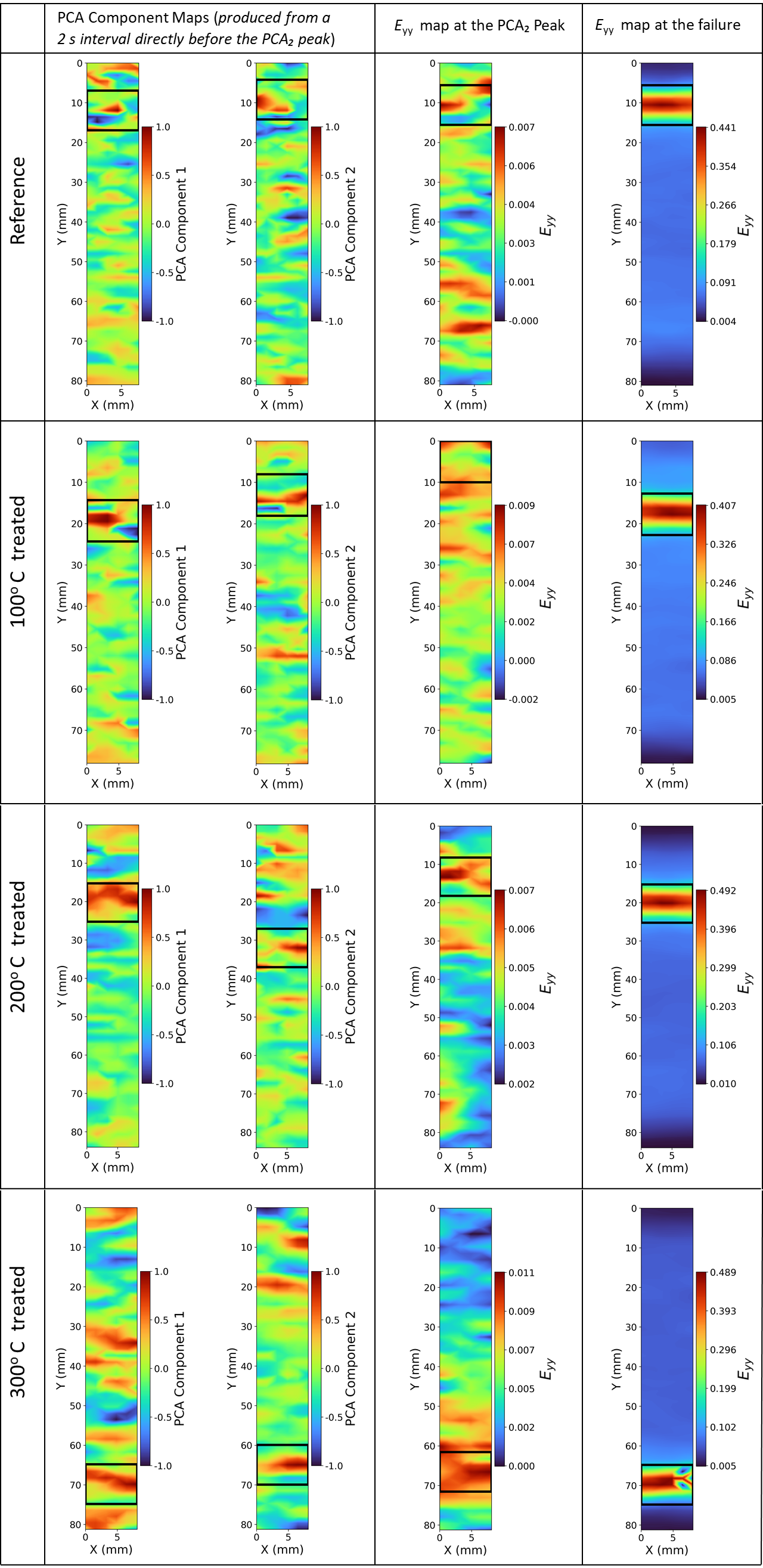}
    \caption{Failure location prediction using PCA for the reference and thermally exposed AA6063-T5 aluminum alloy specimens. The first two columns show the first and second principal component maps generated from 120 consecutive $E_{yy}$ strain maps within the 2~s interval immediately preceding the detected PCA$_2$ peak identified in Figure~\ref{fig:projections}. The third column presents the corresponding $E_{yy}$ strain map at the detected PCA$_2$ peak, while the final column shows the $E_{yy}$ strain map immediately before failure. For visualization, the black rectangular boxes are centered on the maximum value of each map and extend approximately $\pm 5$~mm in the vertical direction. The dimensions of all strain maps are given in millimeters.}
    \label{fig:all_pred}
\end{figure*}

Figure~\ref{fig:all_pred} presents the results of the second stage of the computational framework. After identifying the PCA$_2$ peak shown in Figure~\ref{fig:projections}, 120 consecutive full-field $E_{yy}$ strain maps extracted from the 2~s interval immediately preceding the detected transition were subjected to PCA. This analysis was performed to investigate whether localized deformation features associated with the future failure location could be identified from a high-temporal-resolution sequence of strain maps.

The first two columns of Figure~\ref{fig:all_pred} present the first and second PCA component maps obtained from the selected image sequence. For all investigated conditions, the first principal component map exhibits a distinct localized region highlighted by the rectangular box. Similar localization features are observed in the second principal component map, although these regions are generally less spatially confined and less consistent than those obtained from the first principal component.

The third column of Figure~\ref{fig:all_pred} shows the corresponding $E_{yy}$ strain maps at the engineering strain associated with the detected PCA$_2$ peak. Although localized high-strain regions (hotspots) are already present, the deformation patterns remain spatially distributed, with several competing hotspots visible, making the eventual failure location difficult to identify directly from the raw strain maps. This behavior is particularly evident for the reference, 100$^\circ$C, and 200$^\circ$C conditions.

For the 300$^\circ$C condition, the raw strain map already exhibits a dominant hotspot near the eventual failure location. However, the localized region remains less spatially confined than in the corresponding principal component maps.

The final column presents the corresponding $E_{yy}$ strain maps immediately before failure. For all investigated conditions, the localized regions identified by the first principal component maps coincide closely with the final strain localization patterns. Although the raw strain maps at the detected PCA$_2$ peak occasionally exhibit elevated strain near the eventual failure location, the first principal component maps consistently provide a clearer and more spatially localized indication of the failure-prone region.

\section{Discussion}
\label{sec:discussion}

The present results demonstrate that thermal exposure influences 
both the global mechanical response and the evolution of local 
deformation in AA6063-T5 aluminum alloy. Although all specimens 
originated from the same initial T5 temper condition, subsequent 
thermal exposure modified the stress--strain response and strain 
localization behavior. This response can be related to the 
precipitation--hardened nature of AA6063 and the sensitivity of its 
strengthening precipitates to thermal history \cite{Aydi2015AA6063T5,Khadyko2017AA6063,Khadyko2019DuctilityAA6063,ZEYBEK2026149761,Munoz2020HeatTreatmentAA6063}. Sekhar et 
al.~\cite{Sekhar2019AgingAA6063} showed that the maximum hardness and 
yield strength of artificially aged AA6063 decrease with increasing 
aging temperature and that the overage condition is characterized 
by precipitates substantially coarser than the overage condition. 
Since the samples investigated here were initially in the T5 
temper, precipitation strengthening was already established prior 
to additional thermal exposure. The subsequent reduction in 
strength can therefore reasonably be associated with thermally 
induced evolution of the existing precipitation state. However, as 
no direct microstructural characterization was performed, this 
interpretation is based on the measured mechanical response and the 
established aging behavior of AA6063 rather than direct observation 
of precipitate evolution.

The influence of thermal exposure becomes more apparent when the 
deformation process is examined through DIC. Unlike conventional 
stress--strain curves, which describe the global response of the 
specimen, full-field measurements reveal the spatial redistribution 
of deformation during loading. The $E_{yy}$ strain maps show the 
progressive development of strain localization, consistent with 
previous DIC investigations of heat-treated AA6063 alloys 
\cite{Munoz2020HeatTreatmentAA6063,Khadyko2019DuctilityAA6063}. 
However, direct interpretation of individual strain maps becomes 
difficult during the early stages of plastic deformation because 
several localized high--strain regions may coexist. Consequently, 
the region that subsequently develops into the final localization
zone cannot generally be distinguished reliably from an 
instantaneous strain field alone. Previous work demonstrated that 
PCA applied to 
DIC strain fields can identify the elastic--to--plastic transition 
in Ni--based alloys through the evolution of principal component 
projections \cite{makinen2022detection}. In the present AA6063-T5 
specimens, the engineering strain corresponding to the PCA$_2$ peak 
agrees closely with the yielding strain obtained using the 
conventional 0.2\% offset method. Importantly, the PCA analysis uses 
only the local strain fields and does not incorporate the global 
stress response or a predefined yielding criterion. The 
correspondence therefore indicates that yielding is accompanied by a 
detectable reorganization of the spatial strain distribution. The 
observation of the same PCA$_2$ signature across the investigated 
thermal exposure conditions further indicates that this 
characteristic remains identifiable despite changes in the 
mechanical response produced by thermal treatment.

Beyond the temporal PCA projections, the principal component maps 
provide spatial information about the evolving deformation process. 
Whereas an individual $E_{yy}$ map represents the instantaneous 
strain distribution, the component maps are obtained from the 
temporal sequence of strain fields and therefore represent 
characteristic spatial patterns contributing to the deformation 
evolution. This distinction is particularly important near yielding, 
where several competing hotspots can be present in the raw strain 
maps. The first principal component maps consistently emphasize a 
more localized region that subsequently corresponds closely to the 
final strain localization zone. These regions should not be 
interpreted simply as locations of maximum instantaneous strain; 
rather, they represent spatial features contributing strongly to the 
dominant variance in the evolving strain-field sequence.

This spatial interpretation extends the PCA--DIC approach beyond the 
use of principal component projections for transition detection. A 
related concept was previously demonstrated for creep deformation, 
where component maps obtained using dimensionality reduction 
identified spatial regions associated with subsequent failure 
development \cite{mammadli2025universal}. The present results 
demonstrate that characteristic spatial information can also be 
extracted during monotonic tensile deformation of precipitation-
hardened aluminum. The ability to obtain both temporal and spatial 
information from the same unsupervised analysis is therefore an 
important feature of the proposed methodology: the projections 
provide an indicator of changes in deformation behavior, while the 
component maps reveal the spatial organization of the corresponding 
deformation modes.

An important consideration in the proposed two-stage framework is the selection of the time window used for spatial PCA analysis. In the present implementation, the PCA$_2$ peak was identified using strain fields extending to UTS, and the corresponding transition time was subsequently used to select the preceding high-temporal-resolution strain-field sequence. This procedure was adopted to demonstrate that both yielding detection and identification of regions associated with subsequent failure can be achieved solely from DIC strain fields. However, the spatial PCA analysis does not inherently depend on the first stage, since the analysis window could alternatively be selected using the conventional 0.2\% offset yielding point determined from the stress--strain response. Given the close correspondence between the PCA-detected and conventionally determined yielding transitions, this provides a possible alternative for selecting an early-stage analysis window without requiring strain-field measurements up to UTS. Although the present implementation is retrospective, the results demonstrate that strain fields acquired near yielding contain spatial information associated with the eventual failure location, highlighting the potential for earlier identification of failure-prone regions. The development and validation of an online implementation remain subjects for future investigation.

A further practical advantage of the proposed framework is its computational efficiency. In our previous investigation of creep deformation, DIC strain fields were evaluated at a temporal resolution of 30 frames per second throughout the deformation process, requiring substantial computational resources \cite{mammadli2025universal}. In contrast, the present approach combines an initial low-temporal-resolution analysis with high-temporal-resolution processing restricted to a short interval around the detected transition. This strategy substantially reduces the number of computationally demanding DIC calculations and enabled the analysis to be performed using a single CPU rather than a computer cluster. The resulting reduction in computational requirements may facilitate the application of the methodology in experimental laboratories with limited computing resources.

The present results should be interpreted within the scope of the 
experimental investigation, which was designed to examine the 
applicability of the PCA--based methodology across different thermal 
exposure conditions rather than to establish statistically 
representative material properties. Since one specimen was analyzed 
for each condition, the observed differences in mechanical response 
cannot be used to quantify specimen to specimen variability. 
However, the proposed approach operates independently on the 
temporal sequence of strain fields obtained from each experiment and 
does not require labeled training data or multiple specimens for 
model development. In this context, the investigated thermal 
conditions provide distinct mechanical responses for assessing the 
consistency of the extracted PCA signatures. The identification of 
the characteristic PCA$_2$ peak and the correspondence between early 
spatial component maps and subsequent strain localization across all 
investigated conditions demonstrate the feasibility of the approach 
at the individual-experiment level. Nevertheless, additional 
experiments would be necessary to establish its statistical 
reproducibility and robustness against experimental variability. 
Furthermore, the present investigation is limited to one aluminum 
alloy under monotonic uniaxial tensile loading, while the effects of 
thermal exposure are interpreted based on the measured mechanical 
response without direct microstructural characterization.


Future research should therefore examine the methodology using larger experimental datasets and a broader range of materials and loading conditions. Application to cyclic loading would be particularly relevant, since full-field DIC measurements have been successfully used to monitor strain evolution and localization during low-cycle fatigue \cite{Kopec2024DICFatigue}. Extension to elevated-temperature mechanical testing would also be of interest, as DIC has been demonstrated for full-field deformation measurements during high-temperature tensile loading \cite{Valeri2017HighTemperatureDIC}, while recent reviews have established its broader applicability and identified the principal experimental challenges associated with high-temperature deformation measurements \cite{Yu2021HighTemperatureDIC}. Combining such measurements with complementary microstructural characterization could further clarify the relationship between PCA-extracted deformation patterns and the physical mechanisms governing plastic deformation and strain localization.

\section{Conclusions}
\label{sec:conclusions}

This study investigated the capability of Principal Component 
Analysis (PCA) to extract characteristic temporal and spatial 
deformation signatures from Digital Image Correlation (DIC) 
measurements of AA6063--T5 aluminum alloy subjected to different 
thermal exposure conditions. The primary objective was to assess 
whether PCA-based indicators of plastic deformation and strain 
localization remain identifiable despite changes in mechanical 
response induced by thermal exposure. By combining conventional 
uniaxial tensile testing with full-field strain measurements, the 
study examined two complementary aspects of deformation behavior: 
the identification of the yielding transition and the early spatial 
characterization of regions associated with subsequent strain 
localization.
The temporal analysis demonstrated that the second principal 
component projection exhibits a characteristic peak near the 
yielding transition for all investigated thermal conditions.
The engineering strain corresponding to this peak showed agreement 
with the conventional 0.2\% offset yield criterion, indicating that 
the PCA response remains identifiable across specimens exhibiting 
different mechanical behaviors. Furthermore, the spatial analysis 
revealed that principal component maps extracted from high--temporal-
-resolution strain fields preceding the detected transition contain 
localized features associated with the eventual failure regions. 
These features were generally more clearly distinguishable than the 
competing strain concentrations observed in the corresponding 
instantaneous strain maps, highlighting the additional spatial 
information provided by the PCA representation.

The results demonstrate the feasibility of using PCA to extract 
complementary temporal and spatial information from experimentally 
measured deformation fields at the individual-experiment level. In 
particular, the identification of characteristic yielding signatures 
and early localization patterns across different thermal exposure 
conditions suggests that the approach can provide useful indicators 
of deformation evolution without requiring labeled training data or 
material-specific constitutive models. Although further experimental 
validation is needed to establish statistical reproducibility and 
assess the methodology under more complex loading conditions, the 
present findings support the potential of combining DIC with 
unsupervised dimensionality reduction for investigating plastic 
deformation and identifying failure-prone regions in engineering 
materials.

\section*{Acknowledgements}
We acknowledge support from the European Union Horizon 2020 research 
and innovation program under grant agreement no. 857470 and from the 
European Regional Development Fund via the Foundation for Polish 
Science International Research Agenda PLUS program grant No. MAB PLUS/ 
2018/8. This work was partially supported by the project “Science4Business – Nauka dla biznesu”, funded by the European Union under the European Funds for a Modern Economy 2021–2027 Programme, Priority 2: An Environment Conducive to Innovation, Measure 2.5. The authors also acknowledge the Department of Construction at Ağrı İbrahim Çeçen University (Turkey) for its experimental support.

 \biboptions{sort&compress}
 \bibliographystyle{elsarticle-num} 
 \bibliography{Ref}

\end{document}